# Watch the Gap: Making code more intelligible to users without sacrificing decentralization?

1st Simona Ramos
*Marie Curie Fellow*
*Computer Science Department*
*Universidad Pompeu Fabra*
Barcelona, Spain
simona.ramos@upf.edu

2nd Morshed Mannan
*Max Weber Fellow*
*Robert Schuman Centre*
*European University Institute*
Florence, Italy
https://orcid.org/0000-0001-9700-2173

*Abstract*— The potential for blockchain technology to eliminate the middleman and replace the top-down hierarchical model of governance with a system of distributed cooperation has opened up many new opportunities, as well as dilemmas. Surpassing the level of acceptance by early tech adopters, the market of smart contracts is now moving towards wider acceptance from regular (non-tech) users. For this to happen however, smart contract development will have to overcome certain technical and legal obstacles to bring the code and the user closer. Guided by notions from contract law and consumer protection we highlight the 'information gap' that exists between users (including judges/legal bodies) and the source code. We present a spectrum of low-code to no-code initiatives that aim at bridging this gap, promising the potential of higher regulatory acceptance. Nevertheless, this highlights the 'The Pitfall of the Trustless Dream', because arguably solutions to the information gap tend to make the system more centralized. In this article, we aim to make a practical contribution of relevance to the wide-spread adoption of smart contracts and their legal acceptance by analyzing the evolving practices that bring the user and the code closer.

*Keywords— blockchain, smart contract, information gap*

## I. INTRODUCTION

A smart contract is an automatable and enforceable agreement between two or more parties. The concept of smart contract was first introduced by computer scientist and cryptographer Nick Szabo in the late nineties [1]. Szabo described a smart contract as a 'smart' agreement tool that can automatically execute certain pre-programmed steps. Nevertheless, Szabo didn't argue for the superiority of smart contracts over paper contracts, as he noted that they should not be seen as intelligent tools that can phase out traditional contracts – as traditional contracts are designed to be understood by people and smart contracts by machines.

In 2013, Ethereum's implementation of a virtual machine allowed for snippets of code (smart contracts) to be executed in a decentralized way without third party interference, bringing a whole new spectrum of applications and possibilities. Under this system, parties can coordinate themselves in a peer-to-peer manner, according to a set of protocols and rules incorporated into self-executing smart contract code [2]. This has led some to describe blockchain as a 'trustless' or 'trust-free' technology [3]. However, although 'disintermediation' has been regarded as one of the most innovative traits of blockchain technology (and the smart contracts relying on it), blockchain-based systems are complex socio-technological assemblages. In other words, these systems are made up not only of code, but they also involve large variety of actors operating at different layers [4]. As such, due to economic and game-theoretic incentives propagated across the system, centralization can occur at different layers: in the concentration of mining pools and mining farms, as well as through online exchanges and blockchain explorers. For example, joining a mining pool is common among miners as a way to make mining rewards more predictable due to the increased difficulty level of mining a block [5].

According to [6], blockchains and smart contracts could lead to strengthening trust among colluders, where blockchain can transform non-cooperative games (collusion) into cooperative ones. Hence, blockchain can be considered a type of algorithmically run "confidence machine", in which users rely on the predictability of the technology but which inevitably involve trusting actors (such as developers, miners, wallet service providers, etc.). In other words, blockchains do not eliminate the need for collaboration and trust but provide reliable records and automation for transparent processes that may facilitate cooperation between agents [7].

Theoretically, smart contracts can be created on top of public, decentralized and distributed ledgers, accessible to everyone willing to enter in a contractual relationship of a certain type. However, creating smart contracts requires certain technical knowledge and expertise, where average users are not able to develop nor fully understand a written smart contract code. Hence, for an average user to access smart contracts, he has to resort to a trusted party with sufficient technical expertise. This has ultimately limited the speed of expansion and adoption of smart contracts among the general public and created policy dilemmas among regulators. According to [8], the inability of average consumers to understand and interpret smart contracts in intelligible language has been seen as being contrary to consumer protection and the duty of information. We explain the notion of consumer protection and duty of information in the following section.

Arguably, the smart contract governance model focuses on proof-based automation of pre-stated functions run by the system and puts aside relevant legal rules and practices related to consumer protection and duty of information. In other words, this model focuses on providing function-based information written in a programming language (e.g. Solidity) needed for proper code execution, which may not be understandable to the average user. For example, using Hyperledger Fabric business smart contracts are defined with specific programming-based terminology where function-based queries are executed using transaction logic [9]. Thus, a user would need a trusted party with sufficient technical expertise (e.g., a smart contract developer) to 'translate' business rules and operations in executable programming code. Although worthy of lengthy discussion, our article does not focus on the possibilities and constraints of transposing business processes in an executable code.

In this article, we focus on the issue of 'information gap' that appears when users are not able to understand the smart contract code or be provided with relevant information in an intelligible language. We underline the regulatory concerns this gap raises regarding consumer protection and duty of information. Accounting for a wide-spread adoption of smart contracts, we give examples of several ongoing initiatives that aim at closing the information gap, discussing the limits and opportunities of the prose-to-code paradigm. We maintain that although potentially beneficial for a wider adoption and legal recognition of smart contracts, the proposed solutions introduce a new type of intermediary in the system which ultimately affects the notion of trust and decentralization. Overall, we aim to make a practical contribution of relevance to the wide-spread adoption of smart contracts and bridge the gap that exists between legal and technical research, supporting policy makers in their regulatory decisions concerning smart contracts.

## II. INFORMATION GAP – AN IMPEDIMENT TO THE LEGAL ACCEPTANCE OF SMART CONTRACTS

There is still no consensus on the definition of the term "smart contract" nor a systematic classification of its applications, as this term is still widely discussed among legal and technical experts [10] [11] [12]. In general, the word 'contract' can indicate that the agents involved are fulfilling certain contractual obligations or exercising certain rights and may take control of certain assets within the shared ledger. At present, the application of smart contracts has expanded across several sectors. Some of the noted benefits of smart contracts include a faster, immutable, automated, distributed and more transparent way of creating and executing a contractual relationship.

In response, various legislative bodies and policymakers have initiated discussions over creating an 'innovation' friendly approach to smart contract regulation which could include smart contract as legally binding if certain conditions can be met. For example, Arizona's Governor Doug Ducey signed HB 2417, which clarifies some of the enforceability factors associated with the use of blockchain and smart contracts under Arizona law, in particular with respect to transactions relating to the sale of goods, leases, and documents of title [13]. In June 2017, two other US states - Nevada and Vermont - passed laws concerning blockchain, with the legislation following the regulatory direction enacted in Arizona [14] . In Europe, a statement by the UK Jurisdiction Taskforce reasoned that smart contracts are capable of constituting legally binding contracts provided that the common law requirements for contract formation are satisfied [15].

Contract law is probably the most important private law institution of individual self-determination and autonomy and it has evolved continuously to respond to the emergence of new contract models [16]. The difficulty of transposing abstract (cognitive) concepts into contractual terms had been acknowledged long before the adoption of smart contracts. [17] recognizes the challenges of contract law formation that arise due to limits in human cognition. Limited information and certain behavioral biases can also lead to non-optimal outcomes and efficiency losses in contract formations [18]. In recent decades, contract formation also faces the challenges of digitization, raising issues such as the legal capacity of parties to enter in contracts and the genuineness of (informed) consent. With the ongoing development and evolution of blockchain and smart contract systems, this challenge has taken on new dimensions, such as the publicity of (private) contract terms, the deterministic enforcement of unfair terms and (unless deliberately provided otherwise) the lack of an option to address and amend the inherent incompleteness of contracts.

Regulatory challenges regarding code-to-prose translation and interpretation have been discussed on an EU level [19]. In the European Union, the applicable contract law includes not only respective national contract law but is also strongly influenced by European law. The proposed requirement for information disclosure depends on the type of a contract considered. Both the Directive 2000/31/EC on e-commerce and the Consumer Rights Directive 2011/83/EU focus on the formation of a contract on the internet. They establish pre-contractual obligations for consumer contracts to inform the consumer about relevant facts, which could be interpreted as also containing certain information about security vulnerabilities in a smart contract setting [16]. As noted, failure to provide consumers with information in a clear and comprehensible manner may lead to heavy penalties [19].

Overall, in 'traditional' contract setting, consumers are granted information rights, which alludes to the right of a party to understand the agreement in intelligible language before any contractual arrangement is established. Under Spanish law, for instance, the requirements go even further as consumers who has entered into an agreement that has been drafted by a commercial entity have the right to obtain the terms and conditions of a contract on paper at any time (failure to comply may nullify the contract that would otherwise have formed) [20]. In general, contract terms must be drafted in plain, intelligible language. Contract terms must not only be grammatically clear, but the consumer must be able to understand their economic consequences. This broad understanding of transparency entails that contracts should also provide clear information to agents regarding the potential implications and economic consequences of the contract. In a smart contract setting, the issues that may arise due to misinformation or misunderstanding may entail high economic costs for the contracting parties. [21] point out that 'ex ante information costs to determine all contingencies could make smart contracting overly costly'. As a result of these legal requirements, smart contract developers are faced with a dilemma - even if it is possible to transpose smart contract code into a written paper form, the terms may not be clear, plain and intelligible to the average user.

With increased smart contract adoption, a question has been posed over the need for both lawyers and judges to develop sufficient expertise in understanding smart contract code and the underlying blockchain technology [22]. While some may argue that the role of the lawyer can be analogue to the role of the developer in a smart contract setting, this can be easily debunked. In other words, non-lawyers typically can understand simple short-form agreements as well as many provisions of longer agreements, especially those setting forth business terms, while a non-programmer would be at a total loss to understand basic smart contract [23]. In addition, as it can be seen in Table I the high diversity of programming languages used to code smart contract further increases the complexity of the problem [24]. The authors argue that current programming languages are unsafe in the sense that it is easy to write code that expresses a behavior that is not intended. One reason is that only a few operations are defined by the language itself and that a programmer is allowed to create new functions with arbitrary names. [25] argues that software's plasticity interacts with automation and immediacy to produce consequences that set it apart from both law and physical architecture.

TABLE I

| Platform | Ledger/Consensus | OPCode/Language | Features |
|---|---|---|---|
| Bitcoin | UTXO, PoW | Scrip/Ivy | Linear execution conditions |
| Ethereum | Accounts, PoW | EVM/Solidity | General Purpose computing |
| Neo | Accounts, BFT | NeoVM/C+, Java | Many compilers for high-level language |
| NXT | Accounts, PoS | Temlates/Website Forms | Just parameters, no coding |
| Corda | UTXO, Raft | JCM/ Java, Kotlin | Stateless functions |
| Cardano | UTXO, PoS | IELE / Plutus | Functional programing |
| Tezos | Accounts, PoS | Michelson/Liquity | Formal verification |

As a result, some have argued for the establishment of legal institutions that will help decipher the meaning and intent of the code providing assistance in case of a dispute requiring judication. Decentralized arbitration services providing assistance for disputes (e.g., Kleros) have also appeared as a stepping-stone in bringing technology and law closer. [19] maintains that in addition to future AI systems used for code interpretation, there will be a need for legal-tech experts capable of translating and interpreting smart contracts in natural language. The author notes that these experts will be in high demand and often out of reach for certain parties (e.g., average consumers who cannot afford high fees). This proves not only problematic in private enforcement but also in the absence of litigation [19]. From an economic policy perspective, the idea to establish a system of court-appointed experts to help decipher the meaning and intent of the code may be useful, however it would significantly increase the cost and burden to the legal system. Unlike automated control, ex-post audits are known in general to increase cost for regulators and to be burdensome for businesses and operators [26]. Also as noted in [19], AI and API systems such as GPT3

This work was funded by H2020 ITN Marie S. Curie Action Grant.

and NaturalyCode are still not fully developed, providing not precise code to prose translation.

The idea to use human based oracles (as external entities) to verify the validity of a contract in terms of consumer protection and provision of relevant information has also been suggested [16]. However, in this case, provision of contracts to external entities would still require some sort of precise conversion between code-to-prose. [27] argues that programmers and lawyers should work together to create better smart contracts, while legislators focus on laws to ensure that smart contract code is audited by trusted third parties.

In the following section we discuss the initiatives aiming at closing the information gap that exists between non-technical agents within a predominately technical setting. We show a spectrum of low-code to no-code projects that introduce a new element to the smart contract governance model.

III. CLOSING THE INFORMATION GAP: COULD NO-CODE INITIATIVES SOLVE THE CONSUMER INFORMATION PROBLEM?

*A. Low-Code Initiative*

A 'low-code' platform is usually designed to make it easier for users to become blockchain developers, while in the case of 'no-code' initiatives, users are not required to have developer knowledge to interact with a smart contract. There have been several market initiatives, ranging on a spectrum between low-code and no-code. For the sake of simplicity, in this article we address these initiatives as Smart Contract as a Service (SCaS). On the low-code side, a platform called Settlemint, specializes in low-code 'tool-kits' for building blockchain apps [28]. Via their platform, a user can interact and deploy a smart contract more easily. In other words, the company offers pre-written smart contract code, "zero config" REST APIs, along with zero-config admin UI and dashboard solutions. A similar initiative, SIMBA Chain's platform enables streamlined low-code smart contract deployment [29]. Low-code initiatives help users upgrade their technical knowledge or assist them in the creation and execution of smart contracts via a provision of technical 'tool-kits'. However, low-code solutions do not mitigate the information gap per se.

*B. The No-Code Initiatives*

'No-code' smart contract initiatives aim at assisting nontechnical users in creating and interacting with smart contracts in their natural language. Although a full spectrum of translations between code and prose doesn't exist, these initiatives provide a solid step towards merging the information gap. As individuals and organizations become more interested in automating routine and business processes, bridging this gap can lead to an increase in the adoption of smart contracts and their applicability.

Most no-code initiatives fall under these two categories: template-based and DSL-based. Templates are the base for document generation. This premise can also be used for generating code-based smart contracts. While the user is not per se involved in creating a smart contract code, what he does is filling up a template based legal contract (written in natural language of the contracting party). The template later gets transposed in a smart contract code via a compiler of other similar computing mechanism provided by the SCaS. The template also gets checked to make sure the data filled by the user is correct and will not modify the expected behavior of the code. A solid implementation of prose based template for smart Ccontract is Openlaw that gives the possibility for users to fill a prose based template and consequently generate contract transposed to a smart contract code on the Ethereum blockchain [30]. MyWish is another no-code smart contract platform, where users fill up a template looking document, specifying their requirements which MyWish later deploys via a smart contract [31].

Extending 'template specifications' creates something called Domain-Specific Language or DSL which is a more complex system that can allow for a higher flexibility and variety than a simple pre- certified prose-based template. A DSL can be seen be 'group' of templates that the user can arrange and fill to be able to define a more complex requirement for code. DSL may be customized to the drafting of contracts for a given sector and can be (i) embedded in a general programming language (understandable for a programmer) or (ii) designed as a separate language (more understandable for a lawyer/average user if using a controlled natural language with user-friendly interface). It may also assist validation that the code is faithful to the agreement [32].

One implementation example is Marlowe Run for the Cardano Blockchain. In the Marlowe Run platform, users can select a type of pre-written financial contract templates, fill it up and run it [33]. For more flexibility and options in building smart contracts Cardano created Marlowe Playground. However, Marlowe Playground is designed for users with some technical/developer knowledge. Hence, although DSL could potentially allow for much higher flexibility and creativity around contract development by no-tech users, its current applicability is still limiting and not as simple for users as template-based agreements. Intentional programming is another area of research that could provide flexibility and ease of use in the future.

### C. Closing the gap

The choice of prose-based templates (to be converted as smart contract) is still limiting as it does not offer a comparable variety as original legal prose agreements do. However, with further development and technological advancement this constraint could be mitigated as new initiatives enter the smart contract market. In other words, SCaS initiatives introduce a new intermediary which could potentially allow for building, testing and legally certifying prose-to-code templates, scaling up as more users adopt smart contracts. Introducing this element also increases the certainty regarding the intent of the contract.

A full implementation of a no-code initiatives would bring the user and code closer by closing the information gap and eventually assist legal bodies with code interpretation in case of a dispute. This seemingly effective solution changes the governance of the system by introducing a new centralized intermediary - raising questions about the trust and confidence in the system. We could argue that the non-technical users of 'no-code' smart contracts would become vulnerable to the whims of those operating these initiatives and users would need to take a 'leap of faith' that these initiatives would act in their best interest [4]. This creates a relationship of trust with the 'no-code' smart contract initiative acting as a trusted party. This is arguably a trust relationship as there is an opportunity for betrayal, by the initiative acting in their own interest or by acting negligently. The uncertainty about whether the initiative is truly 'translating' smart contracts into natural language remains. This has broader implications for confidence in the overall blockchain network. According to [4], it is confidence, as opposed to trust or 'trustlessness' which under-girds blockchain networks and this confidence is premised on the predictability and reliability of how the technology functions. In contrast with trust, confidence is a state of expectation that is based on inductive knowledge gained through past experience and there is no opportunity for expectations to be betrayed. This knowledge does not have to be gained through first-hand mastery of the subject-matter, for instance the finer points of programming a smart contract, but instead arises from experience, common knowledge and reliance on expertise. Given the importance of trust in expert systems (i.e., professionals and their organized knowledge) in accrediting or credentialing expertise, we not only see how confidence is linked to trust but also that a loss of trust in a particular expert system (e.g., 'no-code' software developers) can cause an undermining of confidence (e.g., in the overall blockchain network). While the need for trust persist in public and 'permissionless' blockchain networks, the introduction of 'no-code' smart contract initiatives further increases this need to trust—as well as provides a new centralized 'chokepoint' for regulation.

In general, centralization facilitates regulation by states, as in the case of centralized crypto exchanges [34]. Similarly, according to some, 'core developers' of blockchain protocols should be deemed to be fiduciaries and hold fiduciary duties (e.g., duties of care and loyalty) towards users and others who rely on their decision-making, even if liability for software is contractually disclaimed, [35] , thereby opening the prospect of claiming remedies against them should they fail to abide by these duties. In the same vein, SCaS initiatives may also be categorized as fiduciaries of the users who rely on their no-code initiatives, opening up the prospect of liability if there are code errors/bugs and other failures at converting prose-to-code. However, [36] contradicts [35], arguing that the imposition of fiduciary duties misunderstands how public, 'permissionless' blockchains work: even core developers cannot impose their will on network participants, as the latter decide whether they wish to implement a developer's proposal for a modification or upgrade. Moreover, treating developers as fiduciaries could discourage them from participating in what may be considered a socially beneficial project, due to a fear of potential liability - and without them contributing code the system risks disappearing. Admittedly, according to [37], the Internet has shown that a decentralized infrastructure does not necessarily lead to a decentralization of powers within the system. Blockchain networks have also seen the introduction of intermediaries and developed certain 'chokepoints', even if none of these actors can control the operation of the entire network. Nevertheless, the specter of liability, which is gaining greater attention in the context of ongoing EU efforts to regulate crypto-assets, may become a matter of concern for the SCaS eco-system.

[38] discusses the implications of a Smart Contract Templates Framework (STF) to support complex legal agreements for financial instruments, based on standardized templates. Typically, a legal contract would include rights and obligations that accrue to the different parties and are legally enforceable. These are often expressed in complex, context-sensitive, legal prose and may cover not just individual actions but also time-dependent and sequence dependent sets of actions [38]. There may also be overriding obligations on one or more of the parties such that a lack of action could be deemed to be a wrong-performance or non-performance of the contract. That being said, [38] argue that there are two aspects of the semantics of legal contracts being translated into a smart contract code: a) the operational aspects: these are the parts of the contract that can or should be automated, which typically derive from consideration of precise actions to be taken by the parties and therefore are concerned with performing the contract and b) the non-operational aspects: these are the parts of the contract that shouldn't or cannot be automated. In other words, the smart contract code is assumed to be standardized code whose behavior can be controlled by the input of parameters, while some of the values in the template may not have an operational impact and therefore should not be passed to the smart contract code. Hence, transposing legal prose into a smart contract code may require for a clear distinction between operational aspects.

The closer the outlook of a template is to a legal contract the better, as well as the high disclosure of information in natural language and the needlessness of technical knowledge for interaction with the smart contract. The spectrum of initiatives discussed before ranges in terms of these specifications. [24] argue that a natural language specification that can be compiled to smart contract source code could potentially come to be legally enforceable in court. Legal experts have suggested that a solution to this problem may be found in the possibility of creating a type of a template (or set of templates), which could express in a clear manner the legal intent translated into a smart code. The notion of creating standardized 'contract' templates is not new in legal practice. For example, the oneNDA initiative established a singular contract template for NDAs [39]. According to the World Bank, there has been a long tradition of the use of standardized contract agreements for the procurement of goods and services for traditional public works projects [40]. Other initiatives such as Template.net are helping users create their own legal contracts by filling up an already certified template with a legally appropriate contracting structure.

[41] maintains that smart contracts will likely prove suitable for specific industries and sectors. The ability to design a sector/case specific contract template that would be certified, easily reproducible and translatable to a code by a compiler could bring economies of scale and increase adoption of smart contracts. With normal development, smart contracts are usually reviewed by developers to check that there are no bugs or exploits possible. That takes time and is quite costly, sometimes more than the development of the contract. Using a base template or DSL, contracts can be certified to make sure they will produce the correct code on any valid value entered by the user. As such, when the user fills the template, he can generate the code as soon as the template has been completed. This would also make auditing much more cost and time effective and further ease the burden and costs (beyond initial costs) on operators and regulators.

IV. IS DECCENTRALIZATION A DREAM AFTER ALL?

One can argue that while initiatives of this kind solve the information problem in smart contracts as highlighted by relevant law, there is also a governance change as the system moves more towards centralization. In other words, while smart contracts help avoid intermediaries such as lawyers and notaries, a suggested 'no-code' system introduces an intermediary of another kind.

The idea that blockchain-based systems could not fully operate outside of the purview of the law has been discussed by legal academics and experts. Lawrence Lessing argues that even in a smart contract setting the State is always part of the contractual relationship, because the value of a contract comes from its ability to be legally enforceable [22]. Nevertheless, he maintains that with current technological evolution, legal practitioners are not yet fully familiarized nor able to understand or properly interpret code-based contracts. As such, a situation of 'code illiteracy' among judges and other arbitration entities would require establishing a special legal-tech auditing bodies where code intention and interpretation would be checked. **Hence, the question is not whether "no-code" initiatives bring 'centralization' in a 'decentralized world' but rather the real question is: where centralization is placed to account for the information gap– in the hands of public authorities (often ex-post) or private market initiatives (ex-ante).** Overall, the combination of operating initiatives plus legal auditing bodies contributes to an increase in the overall confidence in the system (rather than relying purely on trust and the opening prospects for betrayal).

From an economic point of view, certified intermediaries may be more cost-effective, as the 'check' of code happens ex-ante, while in the case of public legal-tech auditors the correction would be made ex-post, once a problem would have already appeared. [27] argues that when regulating smart contracts, it makes more sense to prevent problems from arising than trying to correct them afterwards. These are not necessarily exclusive approaches, as regulators often opt for complementary solutions [19], especially when implementing a risk-based approach [42]. The idea that attention towards smart contracts and their overall cost will shift from execution to the drafting stage is highlighted by Shadab who argues that parties would have to specify a more detailed range of contingencies and outcomes, before committing themselves to abide by the decisions of a software-driven contract [43]. [44] argues for ex ante focus on code's production. He maintains that through ex-ante guidance of designers' production of technological normativity, it can be ensured that the illegitimate effects toward which computational legalism tends are minimized as far as possible.

Even the most apparently decentralized systems have shown the capacity to produce economically and structurally centralized outcomes [45]. The author maintains that for decentralization to be a reliable concept in formulating future social arrangements and related technologies, it should come with high standards of specificity. This brings into perspective 'the Pitfall of the Trustless Dream' as defined by [37], arguing that despite the promises of decentralization, the governance of the most popular blockchain networks has become highly centralized. A rather optimistic thought would be that, as smart contract market evolves and adoption increases, SCaS initiatives would evolve to achieve a full implementation of prose-to-code translation, where SCaS initiatives would evolve towards being an open source project.

Open source projects are typically organized in a distributed and decentralized manner, where certain factors determine the long-run sustainability of the operations and the community involved [46]. The process of decentralizing SCaS initiatives, if done right, could contribute further to building overall confidence in the system.

## V. CONCLUSION

Despite the benefits supporting smart contract usage (e.g., transparency, automation, immutability, etc.), smart contract adoption by non-tech users is facing an essential obstacle at the intersection between the law and the code. Guided by notions from contract law and consumer protection we highlight the 'information gap' that exists between users (including judges/legal bodies) and the source code. In relation, we present a spectrum of low-code to no-code initiatives (SCaS) that aim at closing this gap and discuss the potential establishment of legal-tech bodies where code intention, translation and interpretation would be checked. We highlight the interconnection between trust and confidence in relation to the introduction of SCaS initiatives in the blockchain eco-system. We not only see how confidence is linked to trust but also that a loss of trust in a particular expert system (e.g., 'no-code' software developers) can cause an undermining of confidence in the overall blockchain network. Lastly, we ponder on the notion of decentralization, arguing that short run centralization is inevitable when trying to close the information gap. Overall, we argue that the combination of operating SCaS initiatives plus legal auditing bodies can contribute to an increase in the overall confidence in the system, and act as complimentary solutions in bringing the user and the code closer.